\documentclass[12pt, preprint]{aastex}

\usepackage{amssymb}
\usepackage{amsmath}
\usepackage{epsfig}
\usepackage{graphicx}
\usepackage{natbib}
\usepackage{multirow}
\usepackage{rotating}
\usepackage{longtable}
\usepackage{verbatim}
\usepackage{url}
\usepackage{hyperref}
\usepackage{caption2}

\begin{document}
\title{
THE EFFECT OF TEMPERATURE EVOLUTION ON THE INTERIOR STRUCTURE OF H${}_{2}$O-RICH PLANETS
}
\author{Li Zeng\altaffilmark{1,a} and Dimitar Sasselov\altaffilmark{1,b}
}
\affil{$^1$Astronomy Department, Harvard University, Cambridge, MA 02138}
\email{$^a$lzeng@cfa.harvard.edu} 
\email{$^b$dsasselov@cfa.harvard.edu}

\begin{abstract}
For most planets in the range of radii from 1 to 4 R$_{\oplus}$, water is a major component of the interior composition. At high pressure H${}_{2}$O can be solid, but for larger planets, like Neptune, the temperature can be too high for this. Mass and age play a role in determining the transition between solid and fluid (and mixed) water-rich super-Earth. We use the latest high-pressure and ultra-high-pressure phase diagrams of H${}_{2}$O, and by comparing them with the interior adiabats of various planet models, the temperature evolution of the planet interior is shown, especially for the state of H${}_{2}$O. It turns out that the bulk of H${}_{2}$O in a planet's interior may exist in various states such as plasma, superionic, ionic, Ice VII, Ice X, etc., depending on the size, age and cooling rate of the planet. Different regions of the mass-radius phase space are also identified to correspond to different planet structures. In general, super-Earth-size planets (isolated or without significant parent star irradiation effects) older than about 3 Gyr would be mostly solid. 

\end{abstract}

\section{INTRODUCTION}

The catalog of observed extrasolar planets now includes more than 1700 members,
and more than 1100 planets have been observed transiting their parent stars~\citep{Rein:2014}. Transiting planets
are particularly valuable for comparative planetology because they provide the planet's radius,
as well as the inclination angle of the planet's orbit with respect to the line of sight. When combined
with the mass determined from radial velocity measurements, the mean
density of the planet can be determined. 

Super-Earths, massive terrestrial exoplanets within the range of $1 M_{\oplus} \lesssim M \lesssim 15 M_{\oplus}$, are now
observed to be relatively common by Doppler shift surveys and transiting observations. The currently discovered super-Earth extrasolar planets suggest diversity among their interior structure and composition -- some being very dense (such as CoRoT-7b~\citep{Leger:2009, Queloz:2009}), and the
others seem much less so (such as GJ 1214b~\citep{Charbonneau:2009}). 
Moreover, among the Super-Earths, it has been speculated that some of them may contain more than 10$\%$ $\sim$ 15$\%$ of H${}_{2}$O by weight, the so-called water planets (or H${}_{2}$O-rich planets). The candidates of those water planets include GJ 1214b, Kepler-22b, Kepler-68b, and Kepler-18b. 
There is no exact definition of H${}_{2}$O-rich planets; however, based on the implication from the planet formation theory, we could propose the range of anywhere between 25$\%$ and 75$\%$ mass fraction of H${}_{2}$O~\citep{Marcus:2010b}. A value of 100$\%$ H${}_{2}$O would be unlikely because silicate, metal and H${}_{2}$O would tend to be mixed in proportions in the protoplanetary nebula. 

The H${}_{2}$O-rich planets could be roughly divided into two types:
\begin{enumerate}
\item planets with their bulk H${}_{2}$O in the solid phase, or solid H${}_{2}$O-rich planets
\item planets with their bulk H${}_{2}$O in the fluid phase (including molecular, ionic, or plasma phases), like Uranus and Neptune in our solar system but smaller, the so-called mini-Neptunes
\end{enumerate}

It is of particular interest to distinguish between the two types. Furthermore, it would be interesting to know if a planet could transition from one type to the other through thermal evolution, such as the heating or cooling of its interior. The division between the two types depends on the phase diagram of H${}_{2}$O and the mass, the bulk composition, and the interior temperature profile of the planets being considered. 
Thus the goal of this paper is to identify regions and boundaries on the mass-radius (M-R) diagram in order to distinguish planets with different phases of H${}_{2}$O within their interior and to understand how the phases of H${}_{2}$O in the interior could change as planets cool through aging. 


The baseline interior structure model is taken from~\cite{Zeng_Sasselov:2013} and~\cite{Zeng_Seager:2008}. Here we simplify a H${}_{2}$O-rich planet to a fully differentiated planet composed of two distinct layers: a MgSiO${}_{3}$ (silicate) core and an H${}_{2}$O mantle. More detailed three-layer model including the metallic iron is available online, ~\url{http://www.astrozeng.com}, as a user-friendly interactive tool. 

\section{H${}_{2}$O PHASE DIAGRAM}
The low-pressure and low-temperature phase diagram of H${}_{2}$O is notorious for its rich and complex structure. At pressures below $\sim 3$GPa and temperatures below $\sim500$ K, the hydrogen bond is mostly responsible for the diversity of phases. However, the high-pressure and high-temperature phases of H${}_{2}$O appear to be similarly complex (the transitions between $\sim$1000 K and 4000 K), as one approaches the plasma phase of H${}_{2}$O and its dissociation at higher temperatures. The interplay between oxygen atom packing and proton mobility seem to account for much of that complexity.

The pressure-temperature plot (Figure~\ref{f1}) shows different H${}_{2}$O phases in the pressure-temperature regime of interest. The phase boundaries are drawn approximately and are obtained either through experiments (summarized by~\citet{Chaplin:2012}) or by first-principle ab initio simulations~\citep{French:2009, Redmer:2011}. The region marked "molecular fluid" lies above the critical point of H${}_{2}$O ($T_c =$ 647 K, $P_c =$ 22 MPa), i.e., supercritical fluid. The transitions between molecular, ionic, and plasma fluids are gradual~\citep{Redmer:2011}. 

Various structures of Ice XI have been postulated to exist at ultra-high pressure beyond Ice X by ab initio simulations. Those structures are yet to be confirmed by experiments~\citep{Hermann:2012, Militzer:2010}. 

The phase above (higher temperature) the previously known solid forms of Ice VII and Ice X is the "superionic" H${}_{2}$O. Superionic solids are known previously for other materials, e.g., PbF$_2$ and AgI. However, for H${}_{2}$O the phase was first predicted theoretically~\citep{Cavazzoni:1999, Goldman:2005} and confirmed later by experiments~\citep{Ji:2011}. In particular, superionic H${}_{2}$O is characterized by a preserved stable oxygen lattice and mobile protons. The ionic conductivity of protons is primarily responsible for the electrical conductivity. The properties of superionic H${}_{2}$O may have remained as an exotic bit of high-pressure physics, if not for the fact that the pressure-temperature profiles of some super-Earths seem to pass close to the triple point between fluid, superionic, and high-pressure ice phases of H${}_{2}$O. 

\section{THERMAL EVOLUTION OF H${}_{2}$O-RICH PLANET}

The thermal evolution models of a 50wt$\%$ MgSiO${}_{3}$-50wt$\%$ H${}_{2}$O planet, of masses 2, 6, 18.5 $M_{\oplus}$, each of age 2, 4.5, and 10 Gyr (billion years), are considered here. The equation of state (EOS) is from~\cite{Zeng_Sasselov:2013}. 
Figure~\ref{f2} illustrates one example of the models. 

\begin{figure}[htbp]
\begin{center}
\includegraphics[scale=0.75]{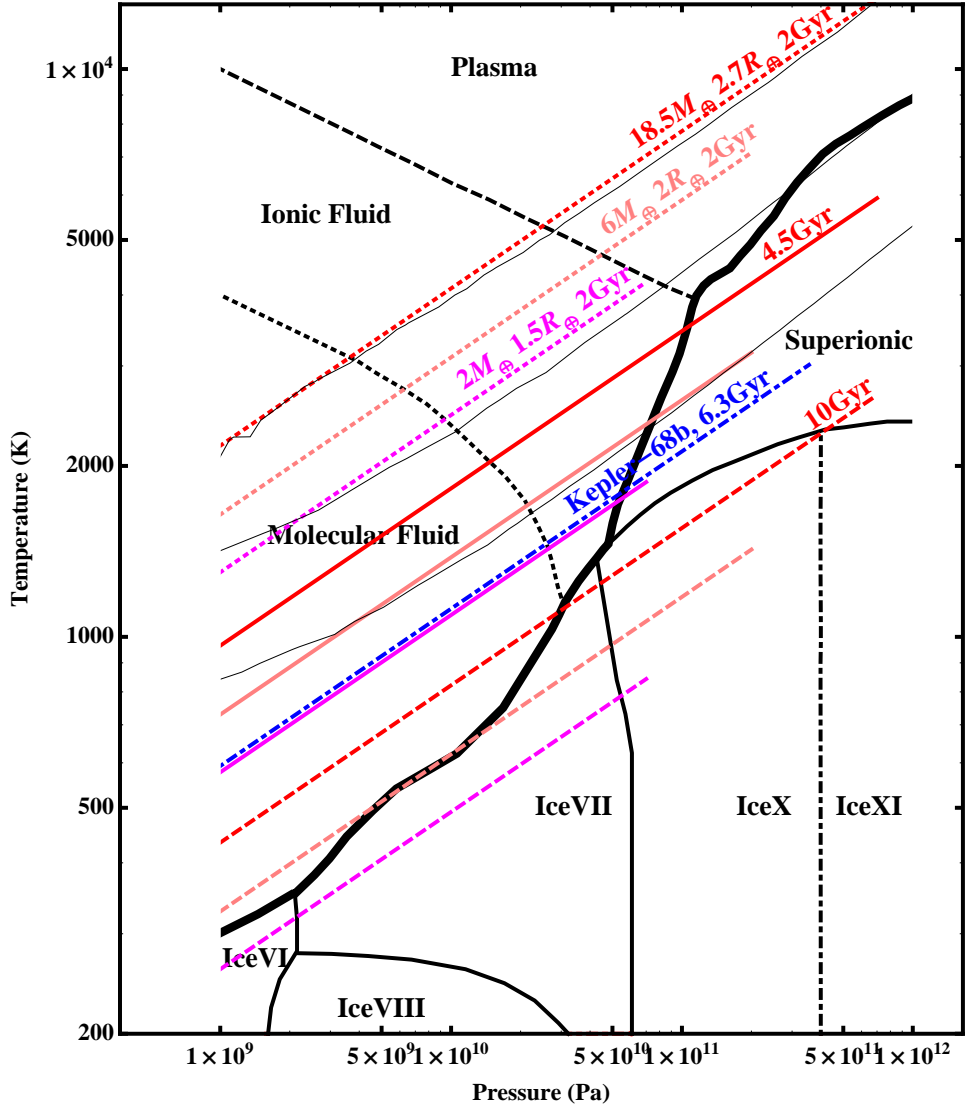}
\end{center}
\caption{Pressure-temperature profiles of H${}_{2}$O-layer of various super-Earth models of different ages, over the H${}_{2}$O phase diagram. The thick black curve is the solid-fluid boundary (melting curve). Three thin black curves are the adiabats calculated from Vazan et al.'s (2013) EOS for comparison. The blue dot-dashed line shows the adiabat for Kepler-68b at the estimated age of 6.3 Gyr~\citep{Kepler-68b:2013}. The nine thermal gradient models as well as the Kepler-68b model are tabulated in Table~\ref{Table1}. The surface pressure of each model is defined as 1 bar ($10^5$ Pa), far beyond the left limit of the diagram. The dotted line indicates the continuous transition from molecular to ionic fluid due to dissociation (more than $20\%$ of the water molecules dissociated), the dashed line indicates the continuous transition from ionic to plasma fluid due to ionization (electronic conductivity $>100 \Omega^{-1}$ cm$^{-1}$) in the dense fluid. The boundary between Ice X and Ice XI is still subject to experimental verification. }
\label{f1}
\end{figure}

\begin{figure}[htbp]
\begin{center}
\includegraphics[scale=0.35]{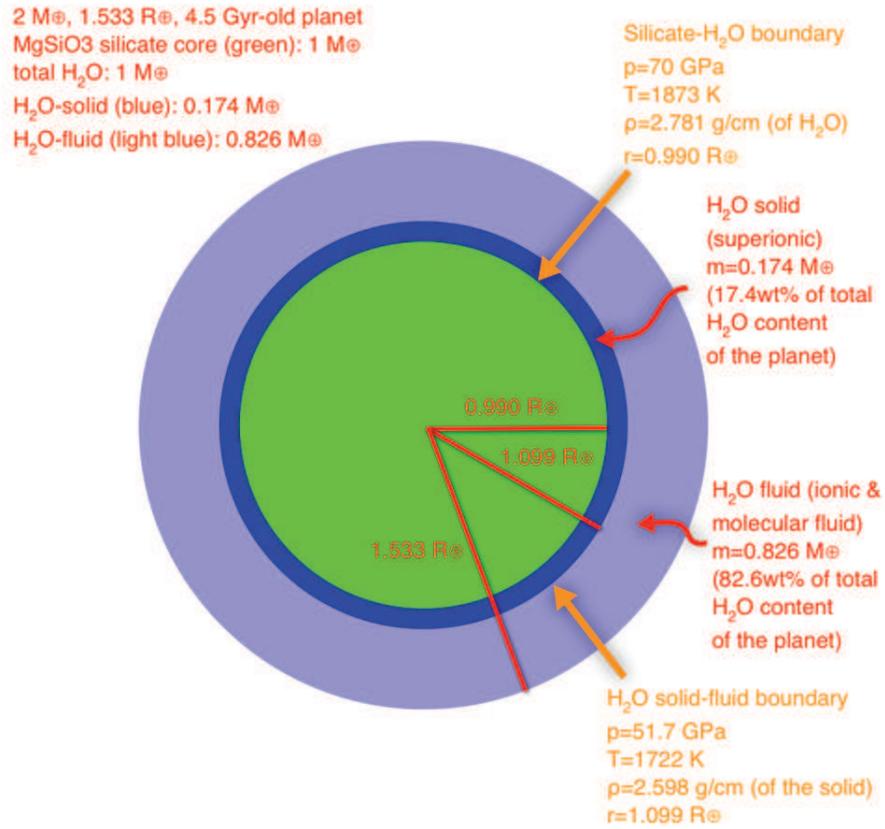}
\end{center}
\caption{Two-layer super-Earth of 2 $M_{\oplus}$ and 1.5 $R_{\oplus}$ at 4.5 Gyr. The interior temperature profile of the H${}_{2}$O-layer of this model is represented by the solid magenta curve in Figure~\ref{f1}.}
\label{f2}
\end{figure}

Figure~\ref{f1} shows the thermal gradients of the models. The three red curves are the models of 18.5 $M_{\oplus}$ and 2.7 $R_{\oplus}$ (large super-Earth, similar to Neptune in terms of mass), of three different ages (2, 4.5 and 10 Gyr). The three pink curves are the models of 6 $M_{\oplus}$ and 2 $R_{\oplus}$ (midsize super-Earth), and the three magenta curves are the models of 2 $M_{\oplus}$ and 1.5 $R_{\oplus}$ (small super-Earth, slightly bigger than Earth). Irradiation by the parent star can have a great effect on the results; most of the super-Earths known today are close to their parent stars. Such planets will stay warm longer. This could increase the length of time they are habitable. For example, the equilibrium temperature of Kepler-68b is estimated to be around 1200 K~\citep{Kepler-68b:2013}, which would have retarded the cooling of the planet from the surface down to about 10 GPa depth at its current estimated age of 6.3 Gyr. 

In order to obtain the initial thermal states to scale from, we have two options. Since we know a lot more details of interior thermal states of solar system planets, compared to exoplanets, it is a good starting point of our model. Most of these H${}_{2}$O-rich planets lie in between Neptune and Earth in terms of their mass and radius; thus we could either scale up from Earth, or scale down from Neptune. Earth is not a H${}_{2}$O-rich planet, so it would make more sense to scale from Neptune. Therefore, we start with Neptune's interior adiabat at the current of age of 4.5 Gyr. We fit an analytical line in log$p$-log$T$ space to Neptune's adiabat~\cite{Redmer:2011}. Then we scale the adiabat to planets of different mass and radius according to essentially their core-mantle boundary temperature T1 and pressure p1, by looking at the similar scaling law of planets in our solar system. Finally, we evolve this scaled adiabat backward or forward to different ages using the rheology law derived in Equation~(\ref{viscosityeq}). In this way, we derive a simple analytical model of planet's interior temperature as a function of its age and pressure: Equation~(\ref{coolingeq}), and Table~\ref{Table1} for a few cases.

Comparing Figure~\ref{f2} to the same model (2 $M_{\oplus}$, 4.5 Gyr) represented by the solid magenta curve in Figure~\ref{f1}, one can see that a small segment of the $P-T$ curve toward the right end (the region near the H${}_{2}$O-silicate boundary) would correspond to a significant mass fraction of H${}_{2}$O inside the planet because the pressure scale is logarithmic in the diagram. 
A simple rule of thumb is that, for the H${}_{2}$O below the depth of $50\%$ p1 (half the H${}_{2}$O-silicate boundary pressure), it contains $\sim40\%$ the total H${}_{2}$O mass, and for the H${}_{2}$O below $10\%$ p1 (one-tenth the H${}_{2}$O-silicate boundary pressure), it contains $>80\%$ H${}_{2}$O mass. For example, the mass of the solid H${}_{2}$O in the 2 $M_{\oplus}$ 4.5 Gyr-old planet is 0.174 $M_{\oplus}$; this is the model illustrated in Figure~\ref{f2}.

\begin{table}[htbp]
\caption{Table of the Pressure-Temperature Profiles of H${}_{2}$O layer\label{Table1}}

\centering
\scalebox{0.64}{
\begin{tabular}{l c c c c c c c |}
\\
\hline\hline
 \multicolumn{7}{c}{Model 1: 2.001 $M_{\oplus}$, 1.533 $R_{\oplus}$}  \\ \hline
$p$ (GPa)\footnotemark[1] & Mass Fraction\footnotemark[2] & Depth\footnotemark[3] (km) & Density\footnotemark[4] (g cm$^{-3}$) & $T$ (K) (2 Gyr)\footnotemark[5] & $T$ (4.5 Gyr) & $T$ (10 Gyr)\\ \hline

0.0001	&	0	&	0	&	0.918	&	101	&	45	&	20.3	\\
1	&	0.0236	&	101	&	1.33	&	1300	&	577	&	260	\\
2	&	0.0466	&	189	&	1.36	&	1570	&	700	&	315	\\
5	&	0.113	&	410	&	1.66	&	2030	&	902	&	406	\\
10	&	0.215	&	740	&	1.83	&	2460	&	1090	&	492	\\
20	&	0.397	&	1310	&	2.09	&	2980	&	1320	&	596	\\
50	&	0.807	&	2700	&	2.58	&	3840	&	1710	&	768	\\
70	&	1	&	3460	&	2.78	&	4210	&	1870	&	843	\\ \hline

 \multicolumn{7}{c}{Model 2: 5.966 $M_{\oplus}$, 2.050 $R_{\oplus}$}\\ \hline
 $p$ (GPa) & Mass Fraction & Depth (km) & Density (g cm$^{-3}$) & $T$ (2 Gyr) & $T$ (4.5 Gyr) & $T$ (10 Gyr)\\ \hline

0.0001	&	0	&	0	&	0.918	&	128	&	56.9	&	25.6	\\
1	&	0.00855	&	60.7	&	1.33	&	1640	&	730	&	328	\\
2	&	0.017	&	114	&	1.36	&	1990	&	884	&	398	\\
5	&	0.0418	&	247	&	1.66	&	2560	&	1140	&	513	\\
10	&	0.0817	&	448	&	1.83	&	3110	&	1380	&	621	\\
20	&	0.157	&	798	&	2.09	&	3760	&	1670	&	753	\\
50	&	0.356	&	1660	&	2.58	&	4850	&	2160	&	970	\\
70	&	0.472	&	2150	&	2.78	&	5330	&	2370	&	1070	\\
100	&	0.625	&	2820	&	3.01	&	5880	&	2610	&	1180	\\
200	&	1	&	4690	&	3.59	&	7120	&	3170	&	1420	\\ \hline

 \multicolumn{7}{c}{Model 3: 18.52 $M_{\oplus}$, 2.687 $R_{\oplus}$}  \\ \hline
 $p$ (GPa) & Mass Fraction & Depth (km) & Density (g cm$^{-3}$) & $T$ (2 Gyr) & $T$ (4.5 Gyr) & $T$ (10 Gyr)\\ \hline

0.0001	&	0	&	0	&	0.918	&	169	&	75.2	&	33.9	\\
1	&	0.00263	&	33.7	&	1.33	&	2170	&	965	&	434	\\
2	&	0.00525	&	63.1	&	1.36	&	2630	&	1170	&	526	\\
5	&	0.013	&	138	&	1.66	&	3390	&	1510	&	678	\\
10	&	0.0258	&	249	&	1.83	&	4110	&	1830	&	822	\\
20	&	0.0506	&	447	&	2.09	&	4980	&	2210	&	996	\\
50	&	0.121	&	933	&	2.58	&	6420	&	2850	&	1280	\\
70	&	0.165	&	1210	&	2.78	&	7040	&	3130	&	1410	\\
100	&	0.227	&	1600	&	3.01	&	7770	&	3460	&	1550	\\
200	&	0.41	&	2690	&	3.59	&	9420	&	4190	&	1880	\\
500	&	0.813	&	5130	&	4.75	&	12100	&	5400	&	2430	\\
700	&	1	&	6400	&	5.33	&	13300	&	5920	&	2670	\\ \hline

 \multicolumn{7}{c}{Kepler-68b Model: 8.3 $M_{\oplus}$, 2.31 $R_{\oplus}$}  \\ \hline
 $p$ (GPa) & Mass ($M_{\oplus}$) & Radius ($R_{\oplus}$) & Density (g cm$^{-3}$) & $T$ (2 Gyr) & $T$ (6.3 Gyr) & $T$ (10 Gyr)\\ \hline 

0.0001&	8.32	&	2.31	&	0.918&	146	&	46.2	&	29.1	\\
1	&	8.29	&	2.3	&	1.33	&	1870	&	592	&	373	\\
2	&	8.26	&	2.29	&	1.36	&	2260	&	718	&	452	\\
5	&	8.17	&	2.28	&	1.66	&	2910	&	925	&	583	\\
10	&	8.03	&	2.25	&	1.83	&	3530	&	1120	&	706	\\
20	&	7.76	&	2.2	&	2.09	&	4280	&	1360	&	856	\\
50	&	7.04	&	2.07	&	2.58	&	5510	&	1750	&	1100	\\
70	&	6.6	&	2	&	2.78	&	6050	&	1920	&	1210	\\
100	&	6	&	1.9	&	3.01	&	6680	&	2120	&	1340	\\
200	&	4.42	&	1.61	&	3.59	&	8100	&	2570	&	1620	\\
300	&	3.3	&	1.37	&	4.04	&	9060	&	2880	&	1810	\\
355	&	2.84	&	1.25	&	4.26(H${}_{2}$O)	&	9490	&	3010	&	1900	\\
355	&	2.84	&	1.25	&	7.06(MgSiO${}_{3}$)	&	 \multicolumn{3}{c}{Core-mantle boundary} \\	
600	&	1.29	&	0.939	&	8.10	&		&		&	\\	
900	&	0.148	&	0.443	&	9.08	&  \multicolumn{3}{c}{MgSiO${}_{3}$ post-perovskite (ppv)} \\	
900	&	0.148	&	0.443	&	9.29	&  \multicolumn{3}{c}{ppv dissociates to MgO and MgSi${}_{2}$O${}_{5}$~\citep{Umemoto:2011}} \\	
1000	&	0	&	0	&	9.61	&  \multicolumn{3}{c}{center of the planet}	\\

\hline
\end{tabular}
}

\end{table}

\footnotetext[1]{Pressure (in giga-Pascal, $10^9$ Pa).}
\footnotetext[2]{Fraction of H${}_{2}$O mass (out of total H${}_{2}$O) above the corresponding pressure/depth.}
\footnotetext[3]{Depth measured from the surface downward in kilometers.}
\footnotetext[4]{Density (in g cm$^{-3}$) at the corresponding pressure/depth.}
\footnotetext[5]{Temperature (in Kelvin) of the age indicated in parentheses.}

The thermal evolution models (the nine thick $P-T$ profiles in Figure~\ref{f1}) are calculated by the following equation:

\begin{equation}
T[\tau,p_{1}][p]=10^{-2.15}\times \frac{4.5~\textrm{Gyr}}{\tau} \times \sqrt{\frac{p_{1}}{1~\textrm{Pa}}} \times \left(\frac{p}{p_{1}}\right)^{0.277}
\label{coolingeq}
\end{equation}

Here $p_{1}$ is the pressure (in Pa) at the H${}_{2}$O-silicate boundary (i.e., the pressure at the bottom of the H${}_{2}$O layer). $\tau$ is the age of the planet in units of billions of years (Gyr); $p$ is an arbitrary pressure within the H${}_{2}$O layer; and $T[\tau,p_{1}][p]$ calculates the corresponding temperature (in Kelvin). The cooling rate can also be influenced by the phase of the H${}_{2}$O in the mantle (different Rayleigh numbers, different convection speeds in different phases, etc.). Equation~(\ref{coolingeq}) assumes a constant cooling rate for all solid phases of H${}_{2}$O. It also assumes that the cooling of the planet is primarily controlled by the viscosity of the solid part of the planet. This assumption is robust as long as the heat transfer mechanism outward is dominated by the temperature-dependent viscosity-driven solid-state convection in the mantle or core. As long as the viscosity has an exponential dependence on temperature, the scaling law is the same. In some cases, mainly in the early evolution, solid H${}_{2}$O part does not yet exist; however, the silicate core of the planet still remains solid. So the assumption here is that the cooling rate of the planet is still controlled by the bottleneck, which is how fast the solid part could convect out heat. 

Phase transitions from fluids to solids are generally exothermic and release energy (latent heat); thus it could also have an influence on the temperature evolution when the H${}_{2}$O in the planet interior transitions from fluid to solid phase, retarding the cooling at the phase transition boundary. However, current experiments could not reach that pressure-temperature regime to measure the latent heat of phase transition yet, and the theoretical calculation has large uncertainties. Therefore, we choose to ignore the latent heat for now.


The temperature gradient in the fluid part of the H${}_{2}$O layer should be adiabatic. Because the viscosity of a fluid is small, any deviation from adiabat would be quickly offset by convection. 
For the solid part of the H${}_{2}$O layer, as pointed out by~\citet{Fu:2010, O'Connell:1980},  the bulk H${}_{2}$O ice mantle would exhibit a whole-mantle convection without partitioning inside, so it is reasonable to approximate the thermal gradient as an adiabat also. 

Equation~(\ref{coolingeq}) represents a family of adiabats, characterized by the same slope in log$P$-log$T$ plot, scaling to different characteristic interior temperatures ($T_i$). 

Equation~(\ref{coolingeq}) is obtained by downscaling the pressure-temperature profile of the interior of Neptune~\citep{Redmer:2011} according to the pressure at the H${}_{2}$O-silicate boundary, and assuming the cooling of the planet is primarily controlled by the rheology (viscosity) of the solid part of the planet (the bottom solid H${}_{2}$O layer, and predominately the silicate core underneath), that is, by how strong the solid part of the planet can convect and transport the heat out. 
Following the argument in~\citet{Turcotte:2002}, assuming an exponential dependence of the viscosity on the inverse of temperature 
\begin{equation}
\mu=\mu_{r}\times \textrm{exp}(\frac{E_a}{RT})
\label{viscosityeq}
\end{equation}
(where $\mu_{r}$ is a constant of proportionality, $E_a$ is the activation energy, and R is the gas constant)
and including the contribution of the radioactive heat sources, one could derive a result showing that the characteristic interior temperature $T_i$ of a planet is, to the first order, inversely proportional to its age.~\citet{Vazan:2013} modeled the evolution of giant and intermediate-mass planets. Three adiabats (thin curves in Figure~\ref{f1}) calculated from their H${}_{2}$O EOS in the region of validity (private communication) are shown to match quite well with our $P-T$ profiles' gradients, confirming the validity of Equation~(\ref{coolingeq}). However, it should be noted that Equation~(\ref{coolingeq}) should only be taken as a qualitative order-of-magnitude estimate because the actual thermal gradient may depend on many other factors, such as different abundance of the radioactive elements in the interior, different initial thermal states, and the surface boundary conditions of the planet. 

The slope of the adiabats are in general shallower than the melting curve, suggesting that for high enough pressure, the adiabat trend would usually intersect the melting curve and result in the high-pressure ice phases or superionic phase usually sitting at the bottom of the fluid phase but not the other way around. 

\section{IMPLICATIONS AND IMPORTANCE OF THE MODELS}

Comparing Equation~(\ref{coolingeq}) to the H${}_{2}$O phase diagram shows that, as a H${}_{2}$O-rich planet ages and cools down, its bulk H${}_{2}$O may undergo phase transition, first from fluid phases to superionic phase, then from superionic phase to high-pressure ices. The timing of these phase transitions would depend on the pressure $p_{1}$ at the bottom of the H${}_{2}$O layer, the initial thermal state of the planet, the abundance of radioactive elements in the interior, and so on. These phase transitions may affect the radius of the planet only slightly, but they may significantly affect the interior convective pattern of the planet and also the global magnetic field of the planet, which results from the dynamo action inside the planet, which in turn depends on the strength of convection, differential rotation, and the electrical conductivity of the convective layer. The existence of the superionic layer is especially favorable for the dynamo action to take place, speculated as probably what is happening in Uranus and Neptune now. As pointed out by~\cite{Stanley:2006} and~\cite{Redmer:2011}, the nondipole magnetic fields of Uranus and Neptune are presumably due to the presence of a conductive superionic H${}_{2}$O shell surrounding the solid core acting as a dynamo. Such a  scenario could similarly exist on other planets that possess such an electrically conductive region (superionic, ionic or plasma phase) of H${}_{2}$O or other species. The implication of the existence of a global magnetic field on the habitability of the planet is also significant, as has been suggested by some people~\citep{Ziegler:2013, Bradley:1994}, and manifested by our own Earth, that the existence of the magnetic field of Earth shortly after its formation is intimately tied to the origin of life on Earth because it shields the harmful UV radiation from the host star and may have something to do with the origin of chirality of biomolecules such as RNA and protein. 

\section{MASSÐRADIUS DIAGRAM AND H${}_{2}$O PHASE REGIONS}

\begin{figure}[htbp]
\begin{center}
\includegraphics[scale=0.31]{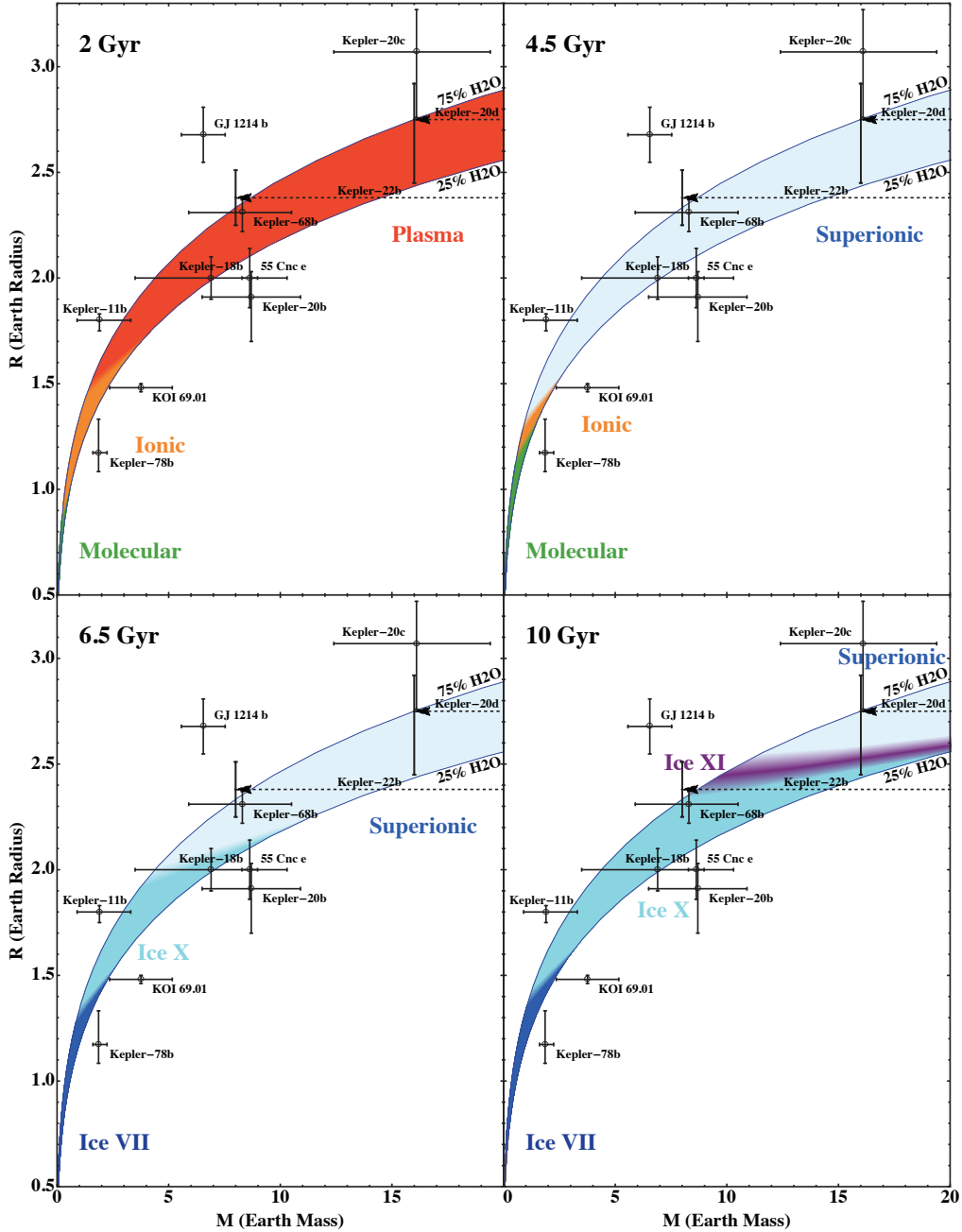}
\end{center}
\caption{ Mass-radius diagram as a function of cooling age corresponding to different phases of H${}_{2}$O near the H${}_{2}$O-silicate boundary, for H${}_{2}$O-silicate planets with H${}_{2}$O mass fraction from 25$\%$ to 75$\%$, of different ages (2, 4.5, 6.5, and 10 Gyr). Exoplanets close to the region of interest are shown, as well as recently discovered KOI 69.01~\citep{Ballard:2013} and Kepler-78b~\citep{Kepler-78b:2013}. }
\label{f3}
\end{figure}

The mass fraction of H${}_{2}$O out of the total planet mass is varied from 25$\%$ to 75$\%$ in the two-layer model, to show the correspondence between different regions of the $M-R$ diagram to different phases of near-bottom H${}_{2}$O for planets of different ages (Figure~\ref{f3}). 

The various colored regions in Figure~\ref{f3} could be compared to the measured masses, radii and ages of observed exoplanets to help us understand the phases of H${}_{2}$O of those planets within this mass range and its implications for planet thermal evolution, convection, magnetic field and habitability. The transport and mixing of volatiles will be different in planets with solid H${}_{2}$O mantle rather than fluid~\citep{Levi:2013}, and that will affect the composition of their atmospheres. For Kepler-68b, there is an accurate age measurement of $6.3\pm1.7$ Gyr from asteroseismology~\citep{Kepler-68b:2013}, which when combined with our model would indicate the presence of solid superionic H${}_{2}$O in its interior.   


One thing to point out is that in our model we have not considered the possible existence of a thick gaseous envelope/atmosphere (such as H/He) that could overlie the H${}_{2}$O layer and increase the observed radii of planets. This gaseous envelope might act as a thermal blanket that would slow the cooling of the planet~\citep{Stevenson:2013}, and instead of interior temperature $T_{i} \sim \tau^{-1}$, it will go as $T_i \sim \tau^{-1/3}$ or even slower. However, because of its low density, it would not increase the interior pressure significantly. We hope to explore this aspect more in future research. 


\section{CONCLUSION} 

We use simple two-layer (silicate-core and H${}_{2}$O-mantle) planet models to understand the thermal evolution of H${}_{2}$O-rich planets. The interior pressure versus temperature profiles of nine specific models are plotted over the H${}_{2}$O phase diagram to show the existence of difference phases of H${}_{2}$O with the thermal evolution of the planets. 

The cooling of a H${}_{2}$O-rich planet results in its bulk H${}_{2}$O content transitioning first from fluid phases to superionic phase, and later from the superionic phase to high-pressure ices. 
These transformations may have a significant effect on the interior convective pattern and also the magnetic field of such a planet, but they may only affect the overall radius slightly. 

Different regions in the mass-radius phase space are identified to correspond to different phases of H${}_{2}$O near the bottom of the H${}_{2}$O layer in a H${}_{2}$O-rich planet, which are usually representative of the bulk H${}_{2}$O in the entire planet (because of the logarithmic pressure scale, a small portion of $P-T$ profile toward the right end would correspond to a considerable amount of H${}_{2}$O by mass). In general, super-Earth-size planets (isolated or without significant parent star irradiation effect) older than about 3 Gyr would be mostly solid. These regions could be compared to observation, to sort the exoplanets into various H${}_{2}$O-rich planet categories, and help us understand the exoplanet population, composition, and interior structure statistically.

The authors are very grateful to Richard O'Connell, Morris Podolak, Jerry Mitrovica, Jeremy Bloxham, Stein Jacobsen, Michail Petaev, Amit Levi, Eric Lopez and David Stevenson for their valuable comments and suggestions and fruitful discussions, and Allona Vazan and Attay Kovetz in particular, for sharing their data of the H${}_{2}$O EOS with us to compare with our model. This work has been supported in part by the Harvard Origins of Life Initiative.

\clearpage


\bibliographystyle{bibstyle1}
\bibliography{mybib}

\end{document}